# Thermal Motions of the E. Coli Glucose-Galactose Binding Protein Studied Using Well-Sampled Semi-Atomistic Simulations

D. J. Cashman, A. Mamonov, D. Bhatt, D.M. Zuckerman

Department of Computational Biology, School of Medicine, University of Pittsburgh

## Abstract

The E. coli glucose-galactose chemosensory receptor is a 309 residue, 32 kDa protein consisting of two distinct structural domains. In this computational study, we studied the protein's thermal fluctuations, including both the large scale interdomain movements that contribute to the receptor's mechanism of action, as well as smaller scale motions, using two different computational methods. We employ extremely fast, "semi-atomistic" Library-Based Monte Carlo (LBMC) simulations, which include all backbone atoms but "implicit" side chains. Our results were compared with previous experiments and an all-atom Langevin dynamics simulation. Both LBMC and Langevin dynamics simulations were performed using both the apo and glucose-bound form of the protein, with LBMC exhibiting significantly larger fluctuations. The LBMC simulations are also in general agreement with the disulfide trapping experiments of Careaga & Falke (*JMB,* 1992; *Biophys. J.,* 1992), which indicate that distant residues in the crystal structure (i.e. beta carbons separated by 10 to 20 angstroms) form spontaneous transient contacts in solution. Our simulations illustrate several possible "mechanisms" (configurational pathways) for these fluctuations. We also observe several discrepancies between our calculations and experiment. Nevertheless, we believe that our semi-atomistic approach could be used to study the fluctuations in other proteins, perhaps for ensemble docking, or other analyses of protein flexibility in virtual screening studies.

## Introduction

Proteins are tiny molecular machines that carry out biochemically relevant functions within all cells. The latter half of the twentieth century has given us over 50,000 individual protein structures, primarily from X-ray crystallography and NMR sources (www.rcsb.org). When analyzing this vast assortment of proteins, it becomes easy to think of the solved structure as the "correct" structure. However, it is important keep in mind that proteins are parts of living organisms, and protein motions are critical to life. These inherent movements play a critical role in locomotion and enzyme catalysis as well as in protein-ligand interactions, and serve as the basis for many biological processes, including, but not limited to, muscle contraction, cellular metabolism, antigen-antibody interactions, gene regulation, and virus assembly.[1-3] Additionally, these movements can be both large-scale, domain movements (e.g. hinge movements), occurring on the order of milliseconds to microseconds, or rapid, small-scale atomic movements, occurring on the nanosecond to picosecond timescale. A recent NMR and modeling analysis of the adenylate kinase system also indicates that large conformational changes may be coupled to catalytic function.[4] These elements of protein conformational changes are also of vital importance in rational drug design, and the development of new methods of incorporating protein flexibility into docking and scoring studies is a topic of much recent research.[5-7]

In this study, we are interested in studying whether our recently developed semi-atomistic simulation approach is suitable for modeling large as well as small scale thermal motions involved in the *Escherichia coli* glucose-galactose binding protein (GGBP). This protein is a 32 kDa globular protein that is organized into two distinct structural domains, each consisting of a $\alpha/\beta$ folding motif. A well-refined X-ray crystal structure exists of this protein (PDB accession code: *2GBP*) which has provided a detailed view of the sugar binding site, located in the cleft between the two primary domains.[8] Some recent studies of this protein have looked at its possible use as a glucose biosensor, which could be used in diabetic patients.[9-11]

A 1992 study by Careaga & Falke experimentally investigated the thermal motions of the alpha-helices on the surface of GGBP by performing a series of disulfide trapping experiments.[12] In their experiments, several residues on two helices of the N-terminal domain, and one residue on the surface of the C-terminal domain (Figure 1), were mutated to cysteine. These cysteine residues were then observed to form disulfide bonds with residue Q26C, in the presence of catalyst, which was quantified by variable mobility in SDS/polyacrylamide gel electrophoresis. These disulfide trapping experiments were used to analyze the thermal motions involved in the protein's hinge movement.[12, 13]

The purpose of this study is to use computational methods to generate an approximate equilibrium ensemble of GGBP based on the published X-ray crystal structure to obtain insight into the mechanism of action by which the experimentally observed fluctuations occur. Are they a result of complete or partial unfolding and refolding of the protein? Or are there rigid-body movements that result in the observed fluctuations? The experimental disulfide trapping measurements provide us with data as to which residues are interacting with each other, and the crystal structure provides us with a clear, yet static, picture of what the protein looks like. We seek to connect these experiments using computational simulations. We therefore performed simulations of both the apo and holo forms of the GGBP protein, using both all-atom molecular dynamics (OPLS-AA force field with the NAMD software package), as well as a coarse-grained Library-Based Monte Carlo (LBMC) simulation method developed in our laboratory.[14] While all-atom molecular dynamics provides the most detailed computational simulation analysis available, it is also computationally demanding, and requires a large amount of single processor CPU time to reach full convergence.[15] In particular, our data suggest that the all-atom simulations are well short of exhibiting motions observed in the Careaga/Falke experiments.[12] Because our coarse-grained LBMC approach uses much less CPU time, it is more likely to exhibit larger-scale motions where residues approach each other close enough for disulfide bond formation to occur.

## Results

This study of the E. Coli D-Glucose/D-Galactose Chemoreceptor protein consisted of four Library-Based Monte Carlo (LBMC) simulations of 3 billion steps in length. One simulation was completed modeling the unbound protein ensemble, and three simulations were completed modeling the glucose-bound protein ensembles. First, the glucose-bound protein was modeled using a "virtual glucose" representation, in which an extra Gō-type interaction was added to those residues that formed either hydrophobic or hydrogen-bonding interactions with glucose in the crystal structure (holo-GGBP "virtual glucose"). A coarse-grained representation of glucose was also added for the other two LBMC models – one representing glucose as a single carbon atom of 1.8 Å in diameter (holo-GGBP 1-C Gō Center), and the other representing glucose as three carbon atoms and of 1.5 Å in diameter (holo-GGBP 3-C Gō Center). Two Langevin dynamics simulations totaling 31 nanoseconds in length (one of the unbound protein and one of the glucose-bound protein) were also performed. Each of the LBMC simulations took approximately 30 days of single CPU time on a 3.6 GHz Intel Xeon system with 2 GB of RAM, while the Langevin dynamics simulations took approximately 141 days of single CPU time on the same system.

### Overall Fluctuations

Figure 2 shows the RMSD vs. Frame for the LBMC simulations (A – D), and RMSD vs. Time for the Langevin dynamics simulations (E, F). In all four LBMC simulations, we observe that the overall trajectory is reasonably stable, with no global unfolding events. However, we do observe some significant fluctuations throughout the trajectory, from 2 Å, up to 5 or 6 Å. Interestingly, the upper limit of the apo-GGBP simulation is markedly higher at certain points (in the range of 5-6 Å), compared to the upper limit of the holo-GGBP ("virtual glucose") and holo-GGBP (1C Go center) simulations, which, for the most part, does not exceed 5 Å (Figure 2. A-C). The upper limit of the holo-GGBP (3-C Go center) simulation does exceed 5.5 Å in two instances (Figure 2D), though the frequency of these events is less than observed with the apo-GGBP LBMC simulation. These events in the apo-GGBP simulation, in which the upper limit falls in the range of 5.5-6 Å, are, in fact, hinge opening events, in which the protein's "mouth" opens up, and then closes again. We also observed hinge opening movements in both of the Langevin dynamics simulations (Figure 2: E, F), although in both the apo-GGBP and holo-GGBP simulations, the protein opens but does not close again. The limited timescale of the all-atom simulations does not permit us to know whether these fluctuations are merely transient.

**Beta Carbon Distances of Key Residues**

To analyze these fluctuations in closer detail, as well as to compare with experimental results, we tracked the beta carbon distance between five residues and Glutamine-26, at intervals of every 200 MC steps throughout the simulations. These five residue pairs: Gln-26/Met-182, Gln-26/Asn-260, Gln-26/Lys-263, Gln-26/Asp-267, and Gln-26/Asp-274, were each mutated to Cysteine residues in separate experiments and observed to form disulfide bonds by Careaga/Falke.[12] Four of these residues are on an alpha helix located immediately adjacent to the alpha helix containing the Glutamine-26 residue (see Figure 1). Methionine-182, on the other hand, is located in the opposite domain of the protein, so tracking the Gln-26/Met-182 $C_\beta$ distance should give us a reasonably good picture of the hinge opening.

Figure 3 shows histograms of the distribution of these beta carbon distances for each residue pair (Figure 3: A-E). First, as expected, we did observe much broader peaks for the histograms generated from the four LBMC simulations compared to the histograms generated from the Langevin dynamics simulations. This is to be expected, as evident from the RMSD plots, we do have a much higher density of large fluctuations in LBMC vs. Langevin. The hinge opening in the Langevin dynamics simulations is clearly evident in Figure 3B for residues Gln-26/Met-182 – instead of two tall, narrow peaks, we see a two broad peaks extending from 26 Å to beyond 40 Å. Unexpectedly, we did not observe any notable difference in the beta carbon distributions between apo-GGBP and holo-GGBP. Table 1 shows the lower and upper ranges of these beta carbon distance distributions, allowing a closer examination of its extremities. We expected that the minimum beta carbon distance observed be lower for the apo-GGBP simulations compared to the holo-GGBP simulations, since the glucose molecule is supposed to stabilize the protein, thereby resulting in decreased intramolecular fluctuations. What was observed in our data is, in fact, the opposite – overall, the lower minimum beta carbon distance observed is in the holo-GGBP simulations, versus the apo-GGBP. Although the beta carbon distance between Gln-26 and Met-182 is consistent with experiment, with a $C_\beta$ distance of 17.5 Å for apo-GGBP and $C_\beta$ distances of 19.7, 18.0, and 17.6 Å, for each of the holo-GGBP LBMC simulations, respectively.

We also observe better consistency with the experimental data overall when comparing the apo-GGBP and holo-GGBP (3-C Gō Center) LBMC simulations.

Our simulations clearly suggest the plausibility of “disulfide-capable” conformations in GGBP. Figure 4.A. shows the relative positions of beta carbons at points spaced every 6 x $10^6$ MC steps apart in the simulation, in which we can see that there is a considerable range of movement. If we cross-reference this with Figure 4: B-D and Table 1, we can observe that there are several instances where the beta carbon distance approaches a very short range, which could make disulfide bond formation possible. Based on a survey of 50 PDB structures in the Protein Data Bank that contain one or more disulfide bonds, we found that the shortest disulfide bond was 3.0 Å and the longest disulfide bond was 4.2 Å, with an average of 3.8 Å. So it’s certainly possible that, for Gln-26/Asn-260, Gln-26/Lys-263, and Gln-26/Asp-267, a disulfide bond could be formed. The $C_\beta$ distances were too great in the LBMC simulations for Gln-26/Met-182 and Gln-26/Asp-274. We also did not observe any distances close to “disulfide-capable” in the Langevin dynamics simulations; see Table 1.

**Rates of Formation of “Disulfide-Capable” Distances**

The rate of formation of “disulfide-capable” beta carbon distances was calculated for the LBMC simulations and compared to the experimental rate constants determined by Careaga/Falke.[12] While time in Monte Carlo simulations is somewhat arbitrary, we can calculate the average number of Monte Carlo steps between instances where the beta carbon distance falls below a certain predetermined threshold (7 Å). Since the experimental rate constant units are in $s^{-1}$, we converted our average number of MC steps to inverse MC steps for better comparison (Table 2). Overall, we observed that the rate of formation of “disulfide-capable” distances was greater for Gln-26/Lys-263, followed by Gln-26/Asp-267, and Gln-26/Asn-260, consistent with the overall trend of the experimental data. We also observe that the rates of formation were also slightly greater in the holo-GGBP LBMC simulations versus the apo-GGBP simulation, which was consistent with our beta carbon distance distribution data, but not with the experimental rate constants. We believe this discrepancy is probably an artifact of the simplicity of our model, which will be improved in subsequent studies.

## Discussion and Conclusion

Overall, the rapid semi-atomistic LBMC simulations provide us with a good picture of the conformational changes and protein fluctuations of the E. Coli D-Glucose/D-Galactose Chemosensory Receptor, and are consistent with previous observations of the open and closed conformations of the protein.[13, 16] Because of the protein's size (309 residues) and the low cost of our simulations (~30 days of single CPU time), we believe that the promise of our semi-atomistic approach has been clearly demonstrated. However, there is certainly room for improvement: discrepancies between our calculations and the experimental data raise cautionary notes.

First, we can clearly see that our LBMC method is a fast, efficient, and robust method that is capable of generating a protein ensemble consisting of several statistically-independent configurations. Our coarse-grained LBMC simulations of 3 billion MC steps all completed in an average of about one month of single processor CPU time, compared to approximately five months of CPU time for the all-atom Langevin dynamics simulations. This represents a significant increase in the rate of observing large fluctuations that lends itself quite well to an investigation of a structurally diverse protein ensemble for use in high-throughput docking calculations, for example.

While we can certainly see reasonably good agreement between our simulations and the experimental data with respect to the plausibility of "disulfide-capable" interactions, which indicates that our observations of the large scale domain movements are good, the discrepancies in the beta carbon distance calculations and the rate calculations are troubling. Perhaps this is evidence that our simplified potential, while good enough to sample the overall conformational space reasonably well, is simply too rough to properly sample the smaller scale fluctuations involved with the presence of glucose. This is one of the reasons why we performed multiple holo-GGBP simulations – first with the "virtual glucose" simulation, then with one carbon representing the glucose molecule, followed by three carbons representing the glucose molecule. Additionally, it was also recently observed that the presence of $Ca^{+2}$ in this protein enhances its thermal stability – neglecting this in our model could have an effect as well.[17] It is also possible that by adding additional chemistry to our coarse-grained model, such as Ramachandran potentials, hydrogen bonding, or residue-specific interactions, we might be able to observe a more realistic sampling of the fluctuations.

Another possibility is that some of the buffers used in the original experiment,[12] could be inadvertently having an effect in promoting disulfide bond formation. For example, one of the catalysts used was $Cu(II)(1,10\text{-phenanthroline})_3$, which is a fairly large molecule (~ 8 Å across), not incorporated into any of our models, and almost certainly not found *in vivo*. It's possibly that there could be some molecular crowding effects caused by this reagent, which is promoting disulfide bond formation between the residues that are far apart, like Gln-26/Met-182 and Gln-26/Asp-274.[18]

In conclusion, while our coarse-grained LBMC method does not do a particularly good job at sampling the small scale protein fluctuations associated with ligand binding to a receptor, its strength lies in its ability to rapidly sample the large scale conformational space of a protein in a short time span. This method should therefore be useful to those interested in high-throughput or ensemble docking calculations involved in rational drug design.

## Methods

We studied the thermal motions of the E. Coli Glucose-Galactose Chemoreceptor using two different computational methods. The first method employs our previously developed Library-Based Monte Carlo (LBMC) with a semi-atomistic protein model.[14] The second method is Langevin dynamics with the OPLS-AA force field[19] and GB/SA solvent.

### Library-Based Monte Carlo

In LBMC, a molecule is divided into non-overlapping fragments and an ensemble of configurations – called a library – is generated in advance for each fragment. During LBMC simulation, fragment configurations in the molecule are swapped with configurations in the libraries. The new state is accepted according to the corresponding acceptance criterion (*vide infra*).

For a molecule divided into $M$ fragments with coordinates denoted by $\vec{r} = \{\vec{r}_1, \ldots, \vec{r}_M\}$ the total potential energy can be decomposed as:

$$U^{tot}\left(\vec{r}_1, \ldots, \vec{r}_M\right) = \sum_{i=1}^{M} U_i^{frag}\left(\vec{r}_i\right) + U^{rest}\left(\vec{r}_1, \ldots, \vec{r}_M\right) \qquad \text{Eq. 1}$$

where $U_i^{frag}$ is the potential energy of individual fragments and $U^{rest}$ represent all other interactions between fragments.

The acceptance criterion of LBMC can be derived using detailed balance conditions. A trial move in LBMC consists of swapping one or several fragment configurations in the molecule with configurations in the corresponding libraries. The old state will be denoted by *o* and the new one by *n*. The acceptance criterion for LBMC swap move can be written as:

$$p_{acc}\left(o \rightarrow n\right) = \min\left[1, \exp\left(-\beta \Delta U^{rest}\right)\right] \qquad \text{Eq. 2}$$

where $\Delta U^{rest} = U^{rest}(n) - U^{rest}(o)$.

Our previous study showed that, for large proteins, LBMC can have a very small acceptance rate.[14] To cope with this problem, we developed a neighbor list trial move, in which new configurations are selected from the neighbor lists instead of the whole library, thereby increasing the acceptance rate. Fragment configurations in the libraries can be classified into the neighbor lists based on some similarity criterion like RMSD, or the some of absolute differences for all bond angles and dihedrals within a fragment. When using a neighbor list trial move, the acceptance criterion should be modified to account for the introduced bias. The acceptance criterion for LBMC swap move from a neighbor list a can be written as:

$$p_{acc}(o \rightarrow n) = \min\left[1, \exp\left(-\beta \Delta U^{rest}\right)\frac{k_i^o}{k_i^n}\right] \quad \text{Eq. 3}$$

where $k_i^o$ and $k_i^n$ is the number of neighbors for the old fragment configuration and the new one, respectively. Note that when the number of neighbors is the same for all configurations in the library, the acceptance criterion of Eq. 2 simplifies to a simpler form of Eq. 3.

**Fragment libraries**

LBMC is flexible with regards to how a molecule can be divided into fragments. In our previous work, we used both peptide-plane and residue based libraries.[14] In this study, we use the same peptide-plane fragments as used in our previous work. Specifically, three types of peptide-planes are employed corresponding to Alanine, Glycine and Proline residues. The peptide-planes span from the alpha-carbon of one residue to the alpha-carbon of the next residue and include all of the backbone degrees of freedom except $\psi$. To allow for the incorporation of Ramachandran potential, the peptide-plane fragments were modified to be conditional on the $\varphi$ dihedral (i.e. to be uniformly distributed in $\varphi$ with a suitable energy correction).

In all of the LBMC simulations, the library size was 2.9 x $10^5$ for Alanine, 2.0 x $10^5$ for Proline, and 3.6 x $10^5$ for Glycine. For all libraries, the neighbor lists were generated to contain 10 configurations. All of the details of neighbor list construction are described in Ref. 10. Libraries of peptide-plane configurations were generated using Langevin dynamics as implemented in the Tinker v. 4.2 software package[20] with the OPLS-UA force field[21] and implicit GB/SA solvent at 298 K.

**Protein Model**

LBMC is flexible in the choice of $\Delta U^{rest}$ which can correspond to standard force field energy terms (e.g., Coulomb and van der Waals interactions) or more approximate interactions such as Gō potential.[22, 23] Following our previous work here, we chose $\Delta U^{rest}$ to correspond to Gō interactions because it can stabilize the native state while at the same time allowing large fluctuations.[14, 24, 25] All the details of Gō interactions are provided in Ref. [22, 23].

**Simulation Details**

The starting structure used for all calculations is the X-ray crystal structure of GGBP with glucose bound (PDB accession coordinates: 2GBP), containing 309 residues.[8] The glucose molecule was removed for the apo-GGBP simulations, and all simulations were run for 3 x $10^9$ MC steps, which followed an equilibration phase of 3 x $10^8$ MC steps. The simulation temperature was chosen slightly below the unfolding temperature based on 13 short simulations of 3 x $10^8$ MC steps, which was determined to be $k_BT/\varepsilon = 0.8$. Frames were saved after every $10^4$ MC steps. Trial moves consisted of swapping three consecutive peptide-planes per step and/or changing the corresponding $\psi$ angles. Due to the large size of this protein, we found that it was optimal to tune the acceptance rate to approximately 20-25% by adjusting two parameters. The first parameter, controlling the ratio of local moves from the neighbor lists to the "global" ones in which configurations are randomly selected from the neighbor lists of neighbor configurations, was set to 10%. The second parameter, controlling the ratio of $\psi$-only moves to the full peptide plane moves, was set to 30%.

The glucose-bound form of the chemoreceptor protein was modeled in three different ways. The first method was a "virtual glucose" method, in which glucose itself wasn't physically added to the model, but an additional Gō type interaction was added to each residue that was observed to make either a hydrogen bond or hydrophobic contact with glucose in the crystal structure (the complete list of residues that additional Gō interactions were added is available in Supplementary Information). These additional interactions are supposed to mimic interactions of the glucose molecule and stabilize the binding site.

The glucose-bound form of the chemoreceptor was also modeled by adding one or three atoms – a coarse-grained representation – of the glucose molecule. The radius of the carbon atom used to represent glucose was 1.8 Å when one atom was used and 1.5 Å when three atoms were used.

The Gō type interactions were added to each atom in the coarse-grained representation of glucose.

The beta carbon distance between Glutamine-26 and Methionine-182, Asparagine-260, Lysine-263, Aspartate-267, and Aspartate-274, was recorded after every 200 MC steps during all of the LBMC simulations. These $C_\beta$ distances would be used to compare our calculated $C_\beta$ distances relative to the crystal structure. We also calculate the rate at which $C_\beta$ distance pairs comes into "disulfide-capable" proximity and compare them to the rates of disulfide bond formation from the Careaga/Falke paper.[12] For a better comparison to the actual rate published by Careaga/Falke,[12] the inverse of this number was recorded. The threshold was selected by surveying 50 disulfide bonds in published X-ray crystal structures from the Protein Data Bank (www.rcsb.org), and determining that the average disulfide bond length was 4 Å. However, none of our LBMC simulations produced a $C_\beta$ distances between residue pairs of 4 Å or less, so a threshold of 7 Å was used. Due to limitations in the software, as well as the fact that it's currently impossible to model actual bond formation of any type, it is highly unlikely that we would have seen a $C_\beta$ distance of 4 Å or less anyway, so if we can use close, "disulfide-capable" distances, these should allow us to compare reasonably well to the disulfide bond formation rates published.

The LBMC source code is available as a free download from our website:
http://www.ccbb.pitt.edu/Faculty/zuckerman/software.html
The modified source code used for the "virtual glucose" and docked glucose calculations is available upon request.

For the Langevin MD simulations, the crystal structure of GGBP (2GBP) was also used as the starting structure. The glucose molecule was removed to simulate apo-GGBP, and it was kept in place to simulate holo-GGBP. The protein was solvated using 11,636 TIP3P water molecules,[26] and simulated using the NAMD software package (version 2.6)[27] with the CHARMM27 all atom force field at 298 K.[28] Frames were saved after every 1 picosecond, and the beta-carbon distance between Glutamine-26 and Methionine-182, Asparagine-260, Lysine-263, Aspartate-267, and Aspartate-274 was calculated from each saved frame.

## Acknowledgements

We would like to thank Bin Zhang, Xin Zhang, Ying Ding, and Andrew Petersen for helpful discussions. Funding was provided by the NIH through grants GM070987 and GM076569, as well as by the NSF through grant MCB-0643456.

## Tables

**Table 1.** Minimum and maximum beta-carbon distances (in Å) from the LBMC and Langevin dynamics trajectories.

| **X-Ray Crystal Value** | **MET-182** 27.8 | | **ASN-260** 12.9 | | **LYS-263** 9.1 | | **ASP-267** 13.2 | | **ASP-274** 19.8 | |
|---|---|---|---|---|---|---|---|---|---|---|
| **Simulation** | **MIN** | **MAX** | **MIN** | **MAX** | **MIN** | **MAX** | **MIN** | **MAX** | **MIN** | **MAX** |
| apo-GGBP MC | 17.5 | 41.7 | 5.7 | 20.8 | 2.2 | 16.7 | 3.0 | 20.6 | 10.2 | 28.7 |
| holo-GGBP MC "virtual glucose" | 19.7 | 36.5 | 5.1 | 23.0 | 0.0 | 18.4 | 3.3 | 20.9 | 6.1 | 27.1 |
| holo-GGBP MC 1C Gō center | 18.0 | 39.4 | 4.6 | 22.2 | 1.0 | 20.1 | 2.0 | 22.6 | 8.7 | 28.3 |
| holo-GGBP MC 3C Gō center | 17.6 | 38.9 | 6.4 | 21.4 | 2.4 | 19.1 | 3.7 | 19.3 | 9.9 | 27.2 |
| apo-GGBP MD | 25.7 | 44.6 | 10.7 | 16.9 | 7.4 | 13.1 | 11.0 | 16.2 | 17.0 | 23.9 |
| holo-GGBP MD | 26.6 | 45.5 | 11.0 | 16.5 | 7.5 | 12.8 | 11.6 | 16.2 | 17.6 | 24.7 |

**Table 2.** Calculated "Rates" for Disulfide-capable interactions in which the $C_\beta$ distance falls below 7 Å.

| **Residue Pair** | **apo-GGBP** | **holo-GGBP** "virtual glucose" | **holo-GGBP** 1-C Gō center | **holo-GGBP** 3-C Gō center | **apo-GGBP** exp. $k_{ss}^{‡}$ | **holo-GGBP** exp. $k_{ss}^{‡}$ |
|---|---|---|---|---|---|---|
| Gln-26 / Met-182 | 0 | 0 | 0 | 0 | $1.40 \times 10^{-2}$ | 0 |
| Gln-26 / Asn-260 | $7.59 \times 10^{-9}$ | $1.70 \times 10^{-8}$ | $1.95 \times 10^{-8}$ | $4.37 \times 10^{-9}$ | $1.60 \times 10^{-2}$ | $1.30 \times 10^{-2}$ |
| Gln-26 / Lys-263 | $9.13 \times 10^{-7}$ | $6.22 \times 10^{-7}$ | $1.06 \times 10^{-6}$ | $8.35 \times 10^{-7}$ | $7.00 \times 10^{-1}$ | $2.10 \times 10^{-1}$ |
| Gln-26 / Asp-267 | $6.98 \times 10^{-8}$ | $1.23 \times 10^{-7}$ | $1.58 \times 10^{-7}$ | $2.00 \times 10^{-7}$ | $7.30 \times 10^{-2}$ | $5.60 \times 10^{-4}$ |
| Gln-26 / Asp-274 | 0 | $2.28 \times 10^{-9}$ | 0 | 0 | $1.50 \times 10^{-2}$ | $1.90 \times 10^{-4}$ |

‡ The Experimental Rate Constant ($k_{ss}$) used is from Careaga, C.L.; Falke, J.J. 1992. *J. Mol. Biol.*, Vol. 226, 1219-35.

## Figure Legends

**Figure 1.** X-ray crystal structure of the E. coli chemoreceptor protein, with several of the key residues of interest highlighted as CPK representations: Red – Gln26, Yellow – Met182, Green – Asn260, Orange – Lys263, Cyan – Asp267, Blue – Asp274.

**Figure 2.** RMSD vs. Frame for Monte Carlo Simulations of GGBP: A. Unbound protein, B. Bound “virtual glucose” protein, C. Bound protein using single carbon atom representation of glucose, D. Bound protein using three carbon atom representation of glucose. RMSD vs. Time for Molecular Dynamics Simulations of GGBP: E. Unbound protein, F. Bound protein.

**Figure 3.** Histograms of the beta-carbon distances from the LBMC and MD simulations for each of the five residue pairs studied: A. Gln-26/Met-182, B. Gln-26/Asn-260, C. Gln-26/Lys-263, D. Gln-26/Asp-267, E. Gln-26/Asp-274. The vertical black line in the center of each graph corresponds to the beta-carbon distance of the 2GBP X-ray crystal structure.

**Figure 4.** A. Shows the positions of $C_\beta$ at frames spaced every 6 x $10^6$ MC steps apart in the apo-GGBP simulation trajectory. B, C, and D show snapshots at various points during the simulation trajectory as the $C_\beta$ distance for Gln-26/Asn-260, Gln-26/Lys-263, or Gln-26/Asp-267 falls below 7 Å. $C_\beta$ are colored as follows: Red – Gln26, Yellow – Met182, Green – Asn260, Orange – Lys263, Cyan – Asp267, Blue – Asp274.

# Figures

## Figure 1

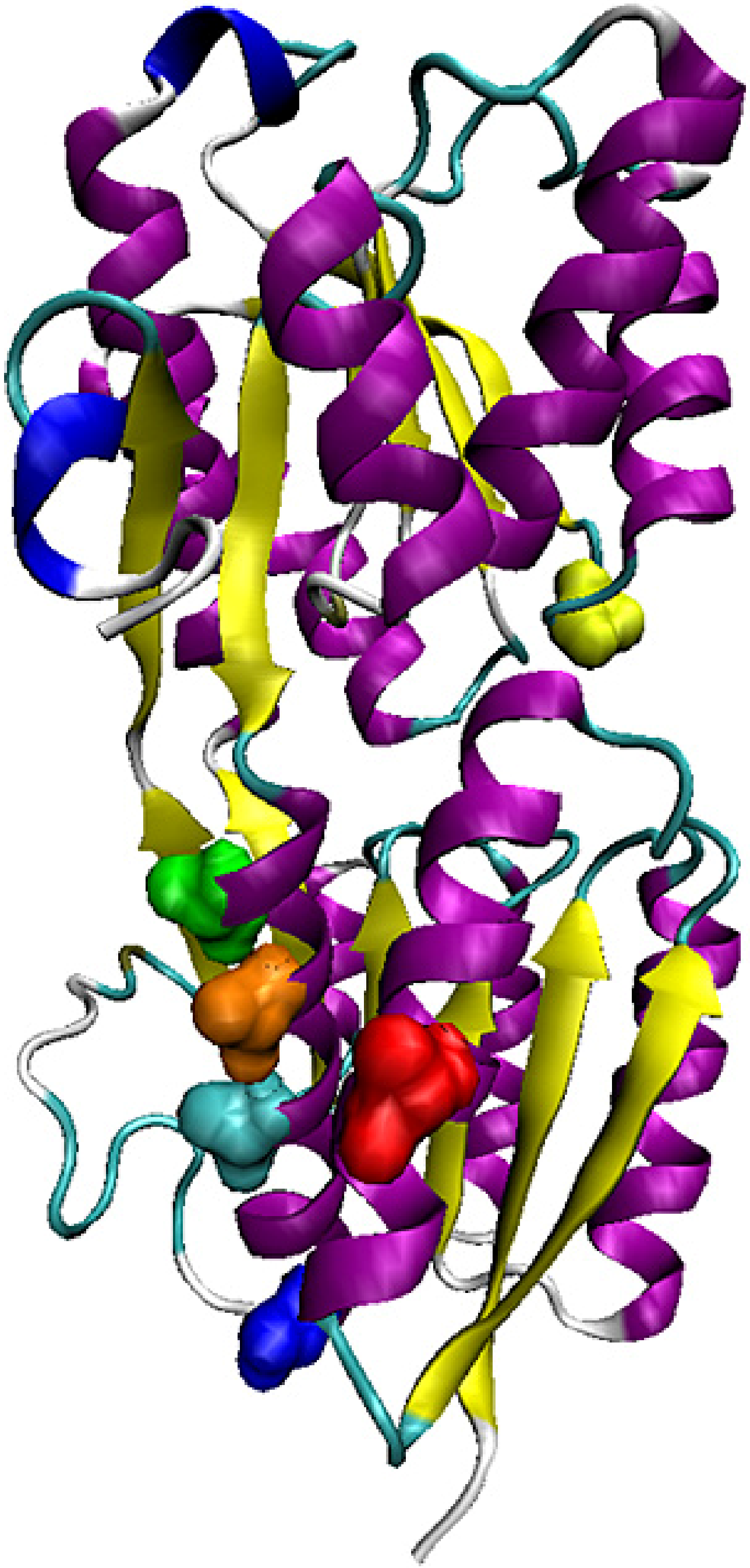

**Figure 2**

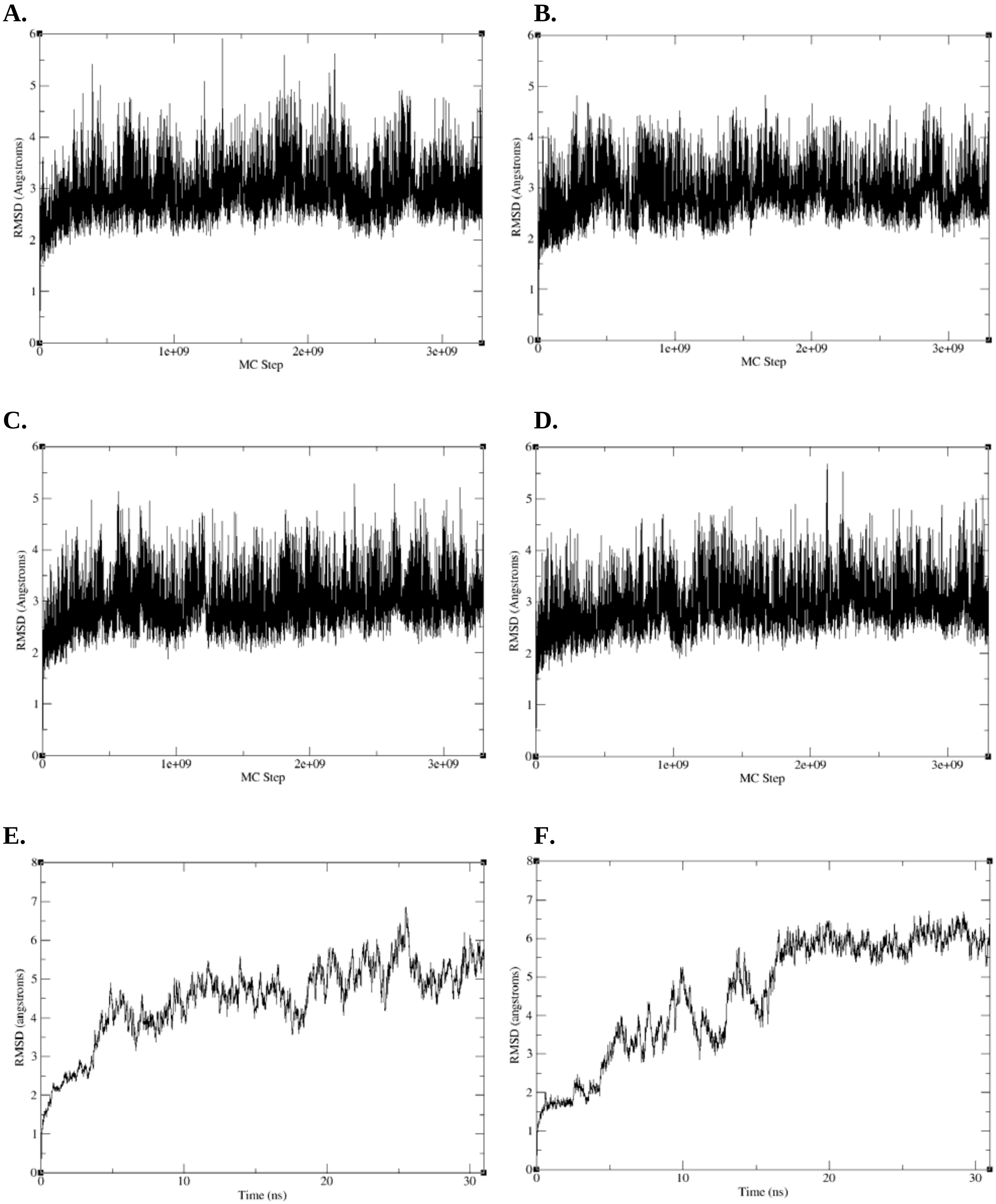

**Figure 3**

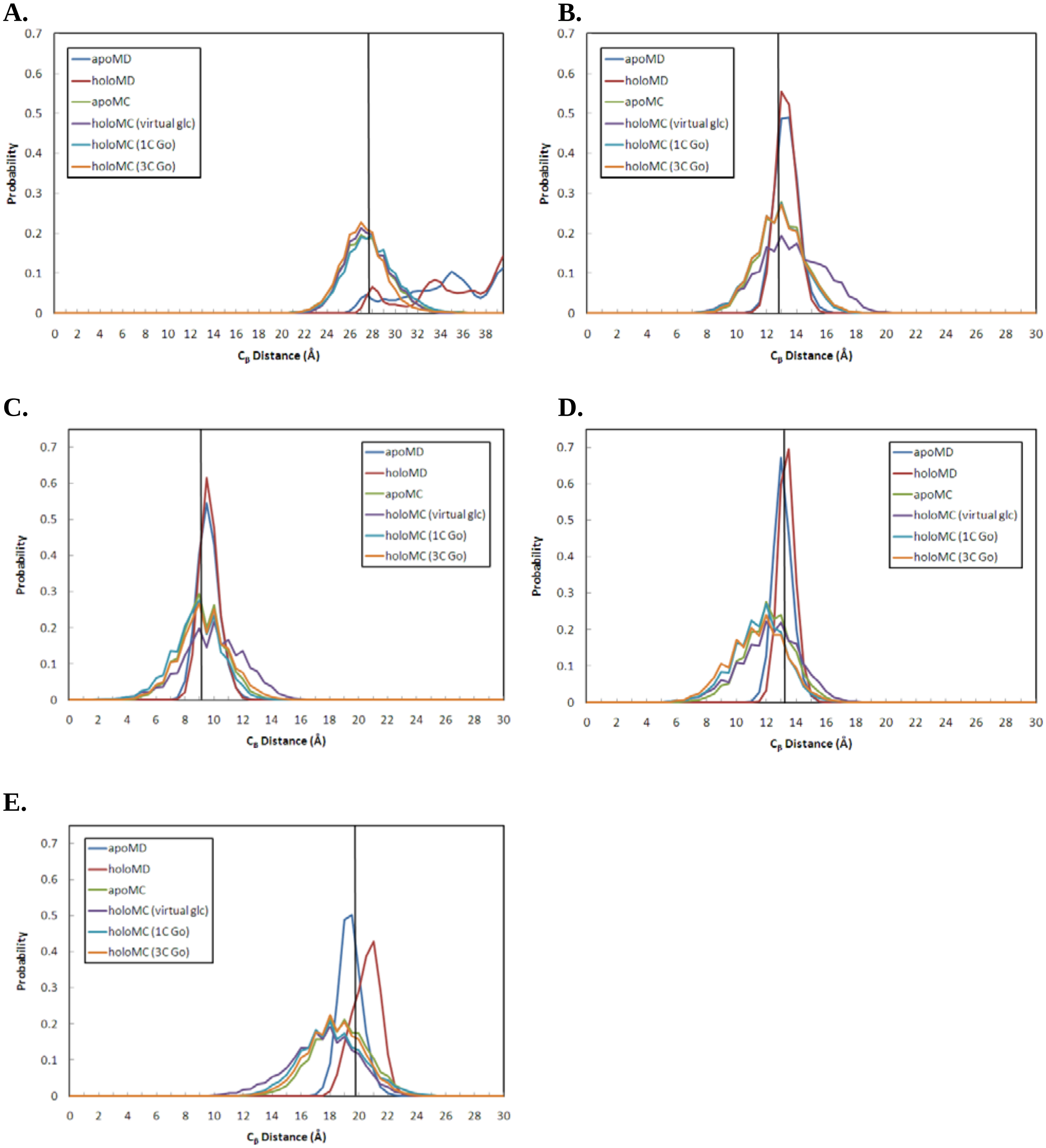
A.
B.
C.
D.
E.
apoMD
holoMD
apoMC
holoMC (virtual glc)
holoMC (1C Go)
holoMC (3C Go)
Probability
$C_\beta$ Distance (Å)
0
0.1
0.2
0.3
0.4
0.5
0.6
0.7
0 2 4 6 8 10 12 14 16 18 20 22 24 26 28 30 32 34 36 38
0 2 4 6 8 10 12 14 16 18 20 22 24 26 28 30

**Figure 4**

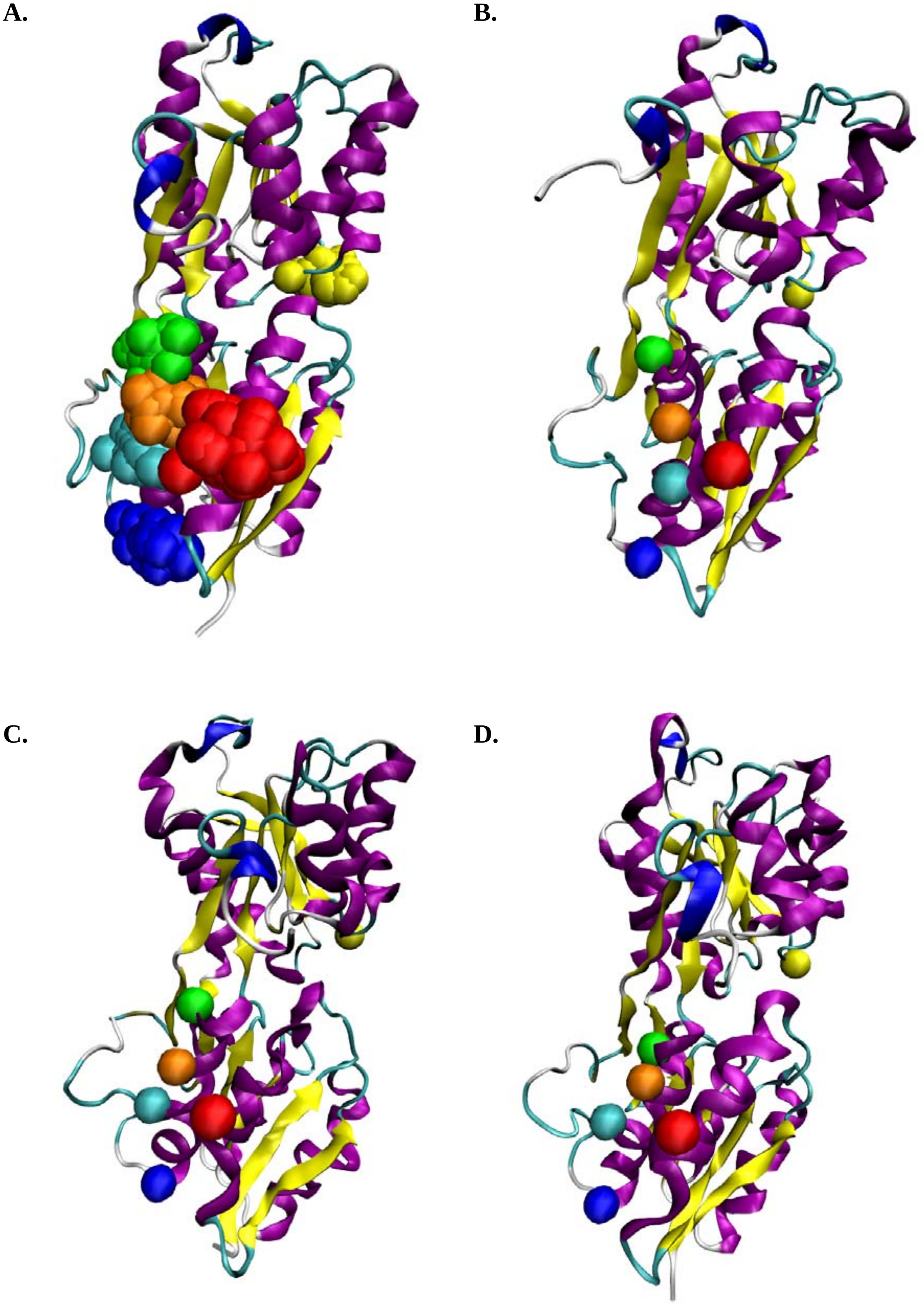